\documentclass[letterpaper,twocolumn,10pt]{ilass}
\def\thepage{}

\usepackage{amssymb}
\usepackage{amsmath}
\usepackage{adjustbox}

\usepackage{ifpdf}
\title{\vspace{-0.25in} {\small \em ILASS-Americas 30th Annual Conference on Liquid
                   Atomization and Spray Systems,
                   Tempe, AZ, May 2019} \newline \newline
        \large {\bf Modeling Drop Deformation Effects in the Euler-Lagrange Prediction of Liquid Jet in Cross Flow} }
\author{\large P. Pakseresht and S.V. Apte\footnote{Corresponding Author: sourabh.apte@oregonstate.edu} \\
        \large School of Mechanical, Industrial and Manufacturing Engineering\\ Oregon State University \\ Corvallis, OR 97331 USA
        \large }

\date{ \normalsize  \centerline{\bf Abstract} \vspace{0.05in}
 \begin{minipage}{6.5in}
 \normalsize Accurate prediction of spray atomization process using an Euler-Lagrange (EL) approach is challenging because of high volume fraction of the liquid phase in dense regimes. This would in reality displace a remarkable portion of the gaseous phase which is commonly ignored in the standard EL approaches. In addition, deformation of droplet due to the interaction of aerodynamic force, surface tension and viscous forces is typically neglected in modeling dense sprays. In this work, to capture the volumetric displacement effects using an EL approach, the spatio-temporal changes in the volume fraction of the gaseous phase are taken into account. This leads to zero-Mach number, variable density equations that give rise to a source term in both momentum and continuity equations. It is shown that the continuity source term increases the velocity and dynamics of the carrier phase close to the nozzle. However, owing to the jet spread and dispersion of droplets, these effects decrease further downstream. In order to quantify the droplet deformation effects, different models are compared together with an experimental data. Different breakup regimes are studied in order to identify the best model for each regime. The shape deformation effect is isolated by performing a single droplet injected into the cross flow with flow conditions similar to the bag-type breakup. A significant deviation in the motion of droplet is observed compared to a case where deformation is neglected.
 \end{minipage} \vspace{-0.25in}
}

\begin{document}

\ifpdf
\DeclareGraphicsExtensions{.pdf, .jpg}
\else
\DeclareGraphicsExtensions{.eps, .jpg}
\fi

\maketitle

\clearpage %This is needed to start a new page.
% Activate page numbers for all pages except the title page. Thus need to reset page counter to 2.
\pagenumbering{arabic}
\setcounter{page}{2}

\section*{Introduction} 
\begin{figure}[h]
\centering
\includegraphics[width=\columnwidth]{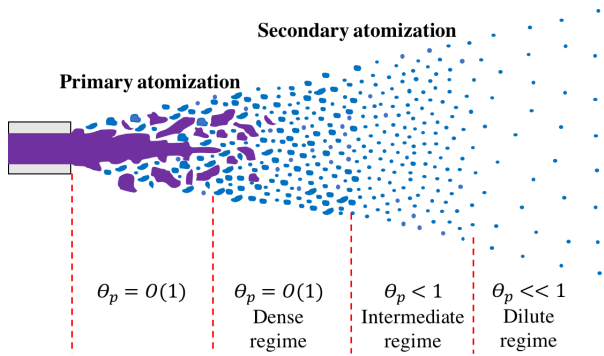}
\caption{Different regimes in a liquid atomization process with illustration of dispersed phase volume fraction.}
\label{fig:spray}
\end{figure}

Liquid spray atomization plays an important role in analyzing the combustion process. A standard modeling approach is to split the process into two steps: primary followed by secondary atomization as shown in Fig. \ref{fig:spray}. Traditionally, the spray dynamics is modeled using an EL point-particle/parcel approach where liquid droplets are assumed subgrid as point droplets and their motion is captured by laws for drag, lift, added mass, and pressure forces. Their effect on the carrier phase is then modeled through two-way coupling of mass, momentum, and energy exchange \cite{Dukowicz1980}. Liquid ``blobs'' with the size of the injector diameter introduced into the combustion chamber undergo atomization based on either deterministic (e.g., \cite{Orourke_1987}) or stochastic breakup models \cite{Apte2003}. In the standard EL point-particle approach, the volume fraction and size of the dispersed phase is assumed small compared to the computational cell. However, this assumption is not strictly applicable to dense spray regions with high void fraction such as those in the primary and the dense regime of secondary atomizations (see Fig. \ref{fig:spray}). This could result in less accurate predictions of such regimes. 

Several works have depicted the importance of accounting for the volume/mass displaced in EL approaches, e.g., \cite{patankar2001a,Snider2001,Ferrante2004,Deen2007,Vanderhoef2008,Apte2008} among others. \cite{Finn2011} and \cite{Cihonski2013} showed that under some conditions, the entrainment of eight small bubbles results in significant levels of vortex distortion when the volumetric displacement effects are accounted for. \cite{Fox2014} showed that new turbulence production terms arise due to correlations between the particle-phase volume fraction and fluid-phase velocity fluctuations. \cite{Capecelatro2014_JFM} observed a strong correlation between the local volume fraction and the granular temperature in the results of fully developed cluster-induced turbulence. \cite{Finn2016} applied this formulation to simulations of natural sand dynamics in the wave bottom boundary layer where overall excellent agreement with experiments was achieved.

Besides, when a droplet is exposed to a high velocity gaseous phase, it undergoes deformation and distortion due to the balance between aerodynamic pressure force, surface tension and viscous dissipation forces. This effect which is typically neglected in the standard EL approaches could ultimately change the breakup process such as breakup time as well as size and velocity of the product drops. However, modeling such effect and coming up with a unique predictive tool for all type of breakup regimes is challenging. \cite{taylor1963} suggested the analogy between an oscillating and distorting droplet and a spring-mass system. In this analogy, the surface tension is analogous to the restoring force of the spring and the aerodynamic pressure force is analogous to the the external force on the mass. \cite{Orourke_1987} added the liquid viscosity as a damping force to this model and modified it as Taylor Analogy Breakup (TAB) model which predicts the breakup process as well. In this spring-dashpot-mass system, forces act on the center of droplet and model its oscillation and deformation. Since droplet is distorted at both north and south equators, therefore the idea of having forces act on the center of droplet was corrected by \cite{Clark1988} in a energy conserved based formulation. In this modified model, each droplet consists of two half drops where forces act on the center of mass of each half. This results in two spring-dashpot-mass system for the given condition. Since Clark's model was linearized and the effect of non-linear deformation particularly for large magnitudes was lost, \cite{Ibrahim_1993} improved their model by more accurately capturing the non-linear effects in large deformations. The three dimensional nature of distorting drop is accounted for by conserving the drop volume instead of area leading to a new Droplet Deformation breakup (DDB) model.  

\cite{Park_2002} improved the original TAB model by modifying the aerodynamic pressure force. This was performed by taking into account the size variation in the projected area of the drop during deformation which was neglected in the original TAB model. \cite{wang2014} developed a model for Bag-Type Breakup (BTB) based on a modified version of the model put forth by \cite{detkovskii1994} wherein kinetic energy of drop is assumed negligible for low Weber number deformations. In their formulation, the expression of deformation was moved to the center of half-drop due to Hill-vortex formation around this point. Similar to \cite{Clark1988} and \cite{Ibrahim_1993}, all forces are applied to the center of mass of half drop. Surface tension is decomposed into two positive and negative parts where the former tends to flatten and the latter restores the drop, respectively. The extension of their model for higher Weber number cases, i.e.,  Multimode Bag (MMB) breakup regime includes the kinetic energy of the droplet \cite{wang2015}. 

\cite{Sor_2015} in the context of ice accretion modified the DDB model by \cite{Ibrahim_1993} by taking into account the accurate calculation of surface tension force. In addition, the instantaneous velocity of the droplet is employed in the deformation model rather than a constant upstream velocity. Furthermore, the center of mass of a half ellipsoid was chosen rather than that of half of sphere. Better predictions on the deformation of a droplet impinging on an airfoil were observed compared to the traditional models, e.g., TAB, Clark's and DDB models. 

In this work, the volumetric displacement effects of deforming droplets in a dense spray are planned for investigations. However, in order to isolate the volumetric displacement effects, Large Eddy Simulation (LES) coupled with Point-Particle (PP) approach modified with spatio-temporal variations in the volume fraction of the carrier phase is employed. For this part, droplets are assumed non-deforming solid particles, and coalescence, breakup and evaporation are all masked to focus only on the displacement effects. Accordingly, a turbulent jet flow laden with a dense loading of solid particles are investigated. Results of this modified LES-PP formulation (volumetric coupling) are compared with those of standard EL point-particle approach (standard coupling) where displacement effects are neglected. For the next step, in order to study the volumetric displacement effects with deforming droplets, different deformation models are investigated by assessing their predictive capabilities for a wide range of Weber numbers and breakup regimes typically observed in sprays. Drop deformation in bag, multimode, transition and shear breakup regimes are all examined to identify a proper model for each regime. These models along with volumetric coupling formulation are deemed to apply to a real atomizing jet in cross flow. However, as a first step in studying the deformation effects, a single liquid droplet injected into a cross flow is examined where the flow parameters are similar to a bag-type breakup condition. It is conjectured that volumetric displacement as well as drop deformation both have to be accounted for in modeling the dense spray regimes.

%%%%%%%%%%%%%%%%%%%%%%%%%%%%%%%%%%%
\section*{Mathematical description}
%%%%%%%%%%%%%%%%%%%%%%%%%%%%%%%%%%%
Carrier phase is captured through solving the governing equations in an Eulerian framework using LES formulation. Motion of liquid droplets is modeled in a Lagrangian framework based on point-particle approach \cite{Maxey1983}. Two phases are coupled primarily by two mechanisms; the displacement of the carrier phase by the volume occupied by the particles and force wise momentum exchange between the phases. The LES volume-averaged governing equations for the carrier phase are as follows 

\begin{equation}
\frac{\partial}{\partial t}\left(\overline{\rho_f\theta_f}\right) + \frac{\partial}{\partial x_j}\left(\overline{\rho_f\theta_f}\widetilde{u}_j\right) = 0
\label{eqn:mass}
\end{equation} 

\begin{equation} 
\begin{split}
&\frac{\partial}{\partial t}\left(\overline{\rho_f\theta_f}\widetilde{u}_i\right) + \frac{\partial}{\partial x_j}\left(\overline{\rho_f\theta_f}\widetilde{u}_i\widetilde{u}_j\right) =  -\frac{\partial \widetilde{P}}{\partial x_i} + \\ 
&\frac{\partial}{\partial x_j} \left(2\overline{\mu_f\theta_f}\widetilde{S}_{ij}\right) -\frac{\partial q^{r,vol}_{ij}}{\partial x_j} +\overline{\rho_f\theta_f}g_i + F_{i,p \rightarrow f}
\label{eqn:momentum}
\end{split}
\end{equation} 

\noindent Here, $\overline{\rho_f\theta_f}$ is the filtered density modified by local volume fraction, $\widetilde{u}$ and $\widetilde{P}$ are the Favre-averaged velocity field and pressure respectively, and $\widetilde{S}_{ij}$ is the Favre-averaged rate of strain. The additional term $q^{r,vol}_{ij}$ in the momentum equation represents the subgrid-scale stress and is modeled using the dynamic Smagorinsky model \cite{moin1991}. As expressed below, rewriting these equations in the form of standard two-way coupling results in extra source terms as $S_{v,cont}$ and $S_{v,mom}$ in the continuity and momentum equations respectively. The former identifies the divergence of velocity due to variation in the local volume fraction whereas the latter gives rise to the volumetric displacement forces. Both source terms are zero in the typical two-way coupling approaches. 

\begin{equation}
\frac{\partial \widetilde{u}_j}{\partial x_j} = S_{v,cont} 
\label{eqn:cont_source}
\end{equation} 

\begin{equation} 
\begin{split}
 &\overline{\rho_f} \left(\frac{\partial \widetilde{u}_i}{\partial t} + \frac{\partial \widetilde{u}_i\widetilde{u}_j}{\partial x_j}\right)  
 =  -\frac{\partial \widetilde{P}}{\partial x_i} + \frac{\partial}{\partial x_j} \left(2\overline{\mu_{f}}\widetilde{S}_{ij}\right) -\\ 
 &\frac{\partial q^{r,2w}_{ij}}{\partial x_j}+\overline{\rho_f}g_i + F_{i,p\rightarrow f} + S_{v,mom}
\label{eqn:momentum_source}
\end{split}
\end{equation}

Throughout this work if the spatio-temporal variations in the local volume fraction is accounted for (i.e., $\theta_f\neq 1$), then volumetric coupling terminology is used whereas the standard two-way coupling (i.e., $\theta_f=1$) is recalled when these effects are neglected. Given the point-particle approach, droplets are tracked using the Newton's second law of motion based on the forces acting on them as 

\begin{equation}
\frac{d{\mathbf x}_p}{dt}= u_p;~\frac{d\mathbf{u}_p}{dt} = \frac{1}{m}_p\left(\mathbf{F}_g  + \mathbf{F}_{p} + \mathbf{F}_{drag} +  
\mathbf{F}_{\rm lift}\right) 
\label{eqn:newton_2} 
\end{equation} 

Different modeling approaches on the droplet deformation are reviewed here. For each model the normalized equations with $y=y/r_o$ and $t=tu_{\infty}/r_o$ are provided. Deformation equation in the TAB model is expressed as follows 

\begin{equation}
   \frac{d^2y}{dt^2} + \frac{5N}{ReK}\frac{dy}{dt} + \frac{8}{WeK}y = \frac{2}{3K}
\end{equation}

\noindent where $N=\mu_l/\mu_g$, $K=\rho_l/\rho_g$, $Re=\rho_gur/\mu_g$ and $We=\rho_gu^2r/\sigma$ are viscosity ratio, density ratio, Reynolds and Weber numbers of drop, respectively. The improved TAB model developed by \cite{Park_2002} in which the aerodynamic force modified during deformation process is obtained as 

\begin{equation}
    \frac{d^2y}{dt^2} + \frac{5N}{ReK}\frac{dy}{dt} + \frac{1}{K}y\left(\frac{8}{We}-2C_F-0.5C_F\right) = \frac{2C_F}{K}
    \label{eqn:Park_model}
\end{equation} 

\noindent where $C_F=4/19$ is chosen such that the critical Weber number, i.e., $We_{crt}=6$ is met. DDB model by \cite{Ibrahim_1993} and its modified version by \cite{Sor_2015} are given below. These two models are different in calculation of surface area as well as the center of mass of half drop. The latter leads to different constant $c$ values of $3\pi/4$ and $8/3$ for DDB and its modified version, respectively.  

\begin{equation}
    \frac{d^2y}{dt^2} + \frac{4N}{ReK}\frac{1}{y^2}\frac{dy}{dt} + \frac{3c}{4KWe} \frac{dA_s}{da} = \frac{3}{8K}c_p
\end{equation}

\noindent where $c_p$ is the pressure coefficient in order to take into account the variations in the gas pressure acting on the droplet surface during deformation. This parameter could be adjusted based on any available experimental data or accurate fully resolved DNS results. $dA_s/da$ for both models is given based on the following expression. Despite the original DDB wherein a simplified version of this parameter was used, its accurate calculation is employed in the modified DDB by \cite{Sor_2015}. 

\begin{equation}
\frac{dA_s}{da} = 
\begin{cases}
\begin{split}
     & 4a - \frac{4}{a^5\epsilon} \ln \left( \frac{1+\epsilon}{1-\epsilon}\right) + \frac{3}{a^{11}\epsilon}[ \frac{2}{\epsilon(1-\epsilon^2)} \\ & -\frac{1}{\epsilon^2} \ln \left( \frac{1+\epsilon}{1-\epsilon}\right) ] \quad \text{Modified DDB}\\
     \end{split} \\
     %\\
    4a(1-2(a)^{-6}) \quad \text{DDB}
    %\end{align}
    \end{cases}
\end{equation}

\noindent where $a=cy$ is the normalized major semi-axis of the half drop and $\epsilon = \sqrt{1-a^{-6}}$. The deformation model in BTB model developed by \cite{wang2014} is expressed as 

\begin{equation}
    \frac{dy}{dt} = \frac{yC_L}{(KN)^{1/3}}\left( \frac{C_d}{2} - \frac{2C_f}{We} \left[ a^{-1} + a^{5} -2a^{-4}\right]\right)
\end{equation}

\noindent where $C_L=C_{d,sph}=0.45$ to account for changes in the pressure on drop surface during deformation from sphere to disk. Comparing with experiments, $C_f=1/600$ was obtained to be the best to close the model \cite{wang2014}. The drag coefficient, $C_d$, is obtained as

\begin{equation}
C_d = 
\begin{cases}
    C_{d,sph} \quad \text{for} \quad (We<10)\\
    2.1 - 13.63/We^{0.95} \quad \text{for} \quad (We\geq10)
\end{cases}
\label{eqn:cd_wang}
\end{equation}

The MMB model by \cite{wang2015} is expressed as 

\begin{equation}
    \begin{split}
    \frac{d^2y}{dt^2} &= \frac{12N}{KRe} [ -\frac{1}{y}\frac{dy}{dt} + \\ & \frac{C_L}{(KN)^{1/3}}\left( \frac{C_d}{2} - \frac{2C_f}{We} \left[ a^{-1} + a^{5} -2a^{-4}\right]\right) ]
    \end{split}
\end{equation}

\noindent where $C_f=0.005$ and $C_d$ is achieved similar to Eq. \ref{eqn:cd_wang} while $C_L$ is obtained as following 

\begin{equation}
    C_L = 
    \begin{cases}
        C_{\mu}(360 - 413.We^{-0.057}) \quad (15<We\leq40)\\
        \begin{split}    
        &C_{\mu}[18.72\exp(5.29\times10^{-3}We) \\
        &+ 0.1125\exp(5.8\times10^{-2}We)] \quad (40<We\leq80)
        \end{split}    
    \end{cases}
\end{equation}

\noindent where 

\begin{equation}
    C_{\mu} = 7.024 \times 10^{-3}Oh^{-4/3}K^{-1/3}
\end{equation}

\noindent and Ohnesorge number, $Oh=\mu_l/\sqrt{\rho_ld_0\sigma}$. Note that in the two last models (BTB and MMB) unlike others, the Weber and Reynolds numbers are calculated based on diameter of drop.

%%%%%%%%%%%%%%%%%%%%%%%%%%%%%%%%%%%
\section*{Results and Discussion}
%%%%%%%%%%%%%%%%%%%%%%%%%%%%%%%%%%%
The numerical approach used in this work has been extensively applied to and validated for different applications \cite{Shams2011,Finn2011,Cihonski2013,Finn2016,Pakseresht2017_asme,pakseresht2017_aps,pakseresht2017_ilass,He2018}. As a first step in modeling dense spray atomization, the volumetric displacement effects of the carrier phase is isolated by masking shape deformation, coalescence, breakup and evaporation of liquid phase by performing LES simulation of a dense particle laden turbulent jet flow. Different particle Stokes numbers are studied to examine their influence on these effects. The studied cases and the corresponding flow parameters are listed in Table \ref{tab:cases_jet}. Detailed explanation and further results can be found in our recent work \cite{pakseresht2019}. For each case, the results of standard and volumetric two-way couplings are compared together. Note that although inter-particle collision is employed for each case (i.e., four-way coupling), in order to solely focus on the particle-fluid interactions, the two-way coupling terminology is utilized. It is imperative to note that to our best of knowledge, no experimental data exist for such dense cases. Thus, only these two formulations are compared in order to investigate the displacement effects of the carrier phase. 

\begin{table}[h]
\begin{center}
\begin{adjustbox}{width=\columnwidth}
\begin{tabular} {lccccc}
\hline
Case & $d_p(\mu m)$ & $Re_{j}$ & S.G. & St & $[\overline{\theta_p}]_{inlet}$ \\
A & 105 & 5712 & 2122.24 & 11.6 & 37.6(\%)  \\ 
B & 105 & 5712 & 7 & 0.0383 & 37.6(\%) \\
\hline
\end{tabular}
\end{adjustbox}
\caption{Flow parameters for different particle-laden turbulent jet cases.}
\label{tab:cases_jet}
\end{center}
\end{table}

\begin{figure}[h]
\centering
\includegraphics[width=\columnwidth]{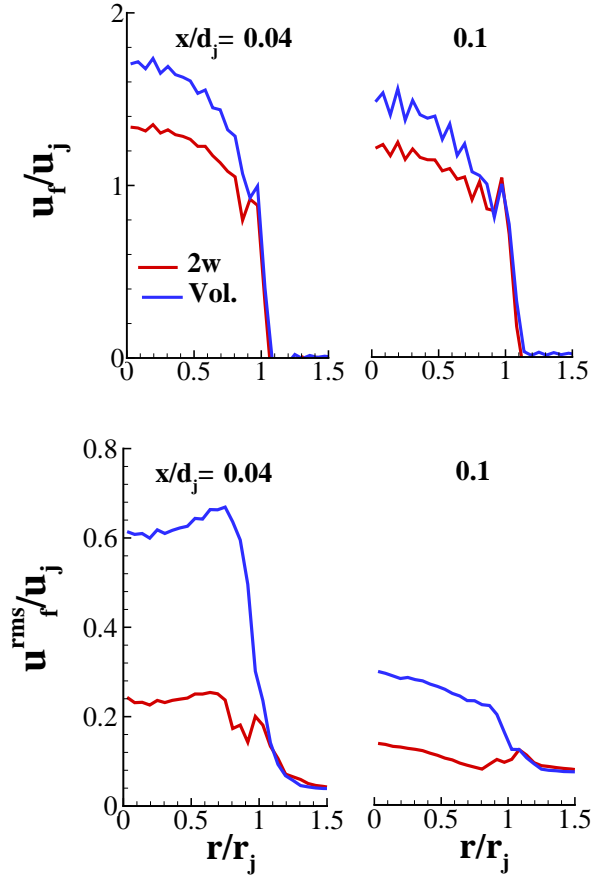}
\vspace{-0.03\textheight}
\caption{Stream-wise mean and r.m.s. velocities of the carrier phase for case A based on the standard and volumetric two-way couplings} 
\label{fig:u}
\end{figure}

Figure \ref{fig:u} shows the results of these two formulations on the mean and r.m.s. velocities of the near field of the jet for case A. As shown, the volumetric coupling predicts higher velocities very close to the nozzle due to the volumetric displacement effects. However, further downstream due to the jet spread and dispersion of particles, the local volume fraction of the carrier phase decreases so do the displacement effects. The increase in the prediction of volumetric coupling is attributed to the continuity source term, $S_{v,cont}$, which drives the higher velocity in the regions with low volume fraction \cite{pakseresht2019}.  

\begin{figure}[!htpb!]
\begin{center}
\includegraphics[width=\columnwidth,keepaspectratio=true,trim={0 0 0 0},clip]{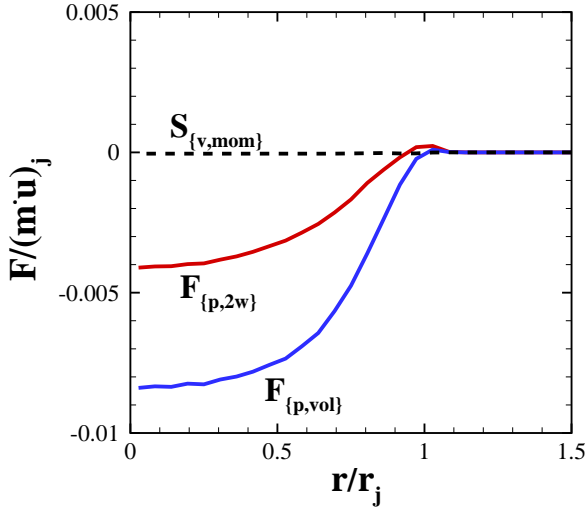}\\
\caption{Radial profile of the normalized time-averaged stream-wise forces in both formulations at the nozzle exit ($x/d_j=0.04$). Shown includes the volumetric displacement force ($S_{v,mom}$) in the volumetric coupling formulation. }
\label{fig:force_fluid}
\end{center}
\end{figure}

As plotted in Fig. \ref{fig:force_fluid}, the contribution of volumetric displacement force ($S_{v,mom}$) in the displacement effects is quite negligible. As shown, the point-particle force in volumetric coupling, $F_{p,vol}$, is predicted almost twice than that of the standard two-way coupling, $F_{p,2w}$. This is due to the higher velocity prediction caused by continuity source term which in turn exerts higher forces on the particles in this formulation. 

\begin{figure}[h]
\centering
\includegraphics[width=\columnwidth,trim={0 0 0 0},clip]{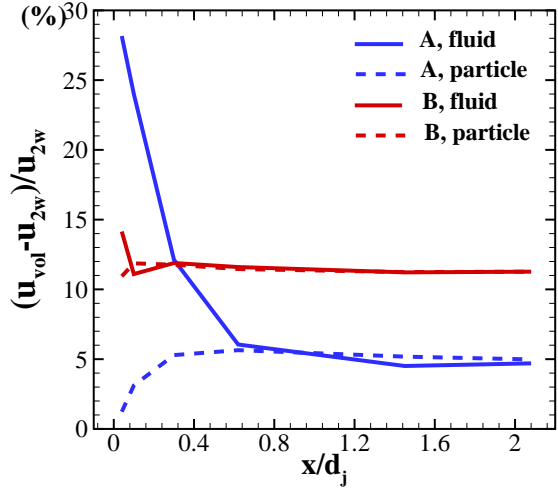}
\vspace{-0.01\textheight}
\caption{Relative increase in the centreline mean velocity prediction of volumetric coupling for both phases of cases A and B.} 
\label{fig:stokes_error}
\end{figure}

The influence of particle Stokes number on the displacement effects is illustrated in Fig. \ref{fig:stokes_error} by looking at the percentage difference on the results of these two formulations. As depicted, decreasing the Stokes number increases the voluemtric displacement effects further downstream. In addition, the dispersed phase gets more affected by these effects. This is attributed due to the fact that particles with lower relaxation time absorb changes in the background flow and react more rapidly to the displacement effects. 

We observed that these effects become important when the inlet average volume loading of the jet is greater than 5\% \cite{pakseresht2019}. For this region, the standard two-way coupling approach is conjectured to be insufficient in order to accurately capture the particle-turbulence interactions. Therefore, for a real atomizing spray where the local volume fraction in the dense regime is on the order of unity, $\theta_p \sim O(1)$, the volumetric displacement effects would be more remarkable, and one needs to account for them. 

As the next step, the deformation effect on the spray characteristics is investigated. It is widely observed that in a spray atomization process, depending on the Weber and Ohnesorge numbers, droplets undergo different phases in terms of deformation and breakup \cite{Krzeczkowski_1980,Hsiang_1992}. For $We<1$, no deformation occurs while drops experience non-oscillatory or oscillatory deformation for $1<We<10$. Increasing Weber number further, would result in more distortion which in turn depending on Weber number, one of the bag, multimode, transition or shear breakup takes place \cite{Hsiang_1992}. Moreover, it is observed that deformation in each breakup regime is quite different \cite{Krzeczkowski_1980}. 

There have been several models predicting the deformation, yet having a model being capable for all regimes is challenging. In this part, the capability of all available models are compared together against the experimental data of \cite{Krzeczkowski_1980} in order to identify the best possible model for each breakup regime. Bag, multimode bag, transition and shear breakup regimes corresponding to the experiment are listed in Table \ref{tab:cases_drop}. 

\begin{table}[h]
\begin{center}
\begin{adjustbox}{width=\columnwidth}
\begin{tabular} {lccccc}
\hline
Case & $Re$ & $We$ & $Oh$ & $N$ & $K$  \\
Bag & 3323.16 & 13.5 & $1.88\times10^{-3}$ & 47.9 & 789 \\ 
Mult. bag & 5161.93 & 18 & $1.4\times10^{-3}$ & 47.9 & 789 \\ 
Transition & 8794.4 & 52.6 & $1.4\times10^{-3}$ & 47.9 & 789 \\
Shear & 12235.69 & 101 & $1.4\times10^{-3}$ & 47.9 & 789 \\
\hline
\end{tabular}
\end{adjustbox}
\caption{Different breakup regimes based on the experimental work of \cite{Krzeczkowski_1980}.}
\label{tab:cases_drop}
\end{center}
\end{table}

\begin{figure}[h]
\centering
\includegraphics[width=\columnwidth]{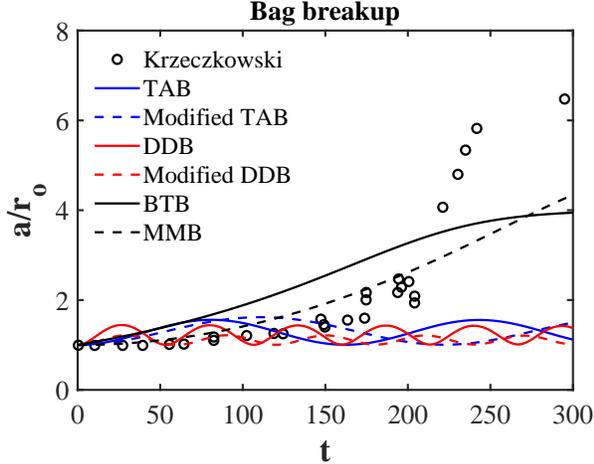}
\vspace{-0.025\textheight}
\caption{Drop deformation in bag breakup regime based on different models compared to the experiment} 
\label{fig:krz_bag1}
\end{figure}

\begin{figure}[h]
\centering
\includegraphics[width=\columnwidth]{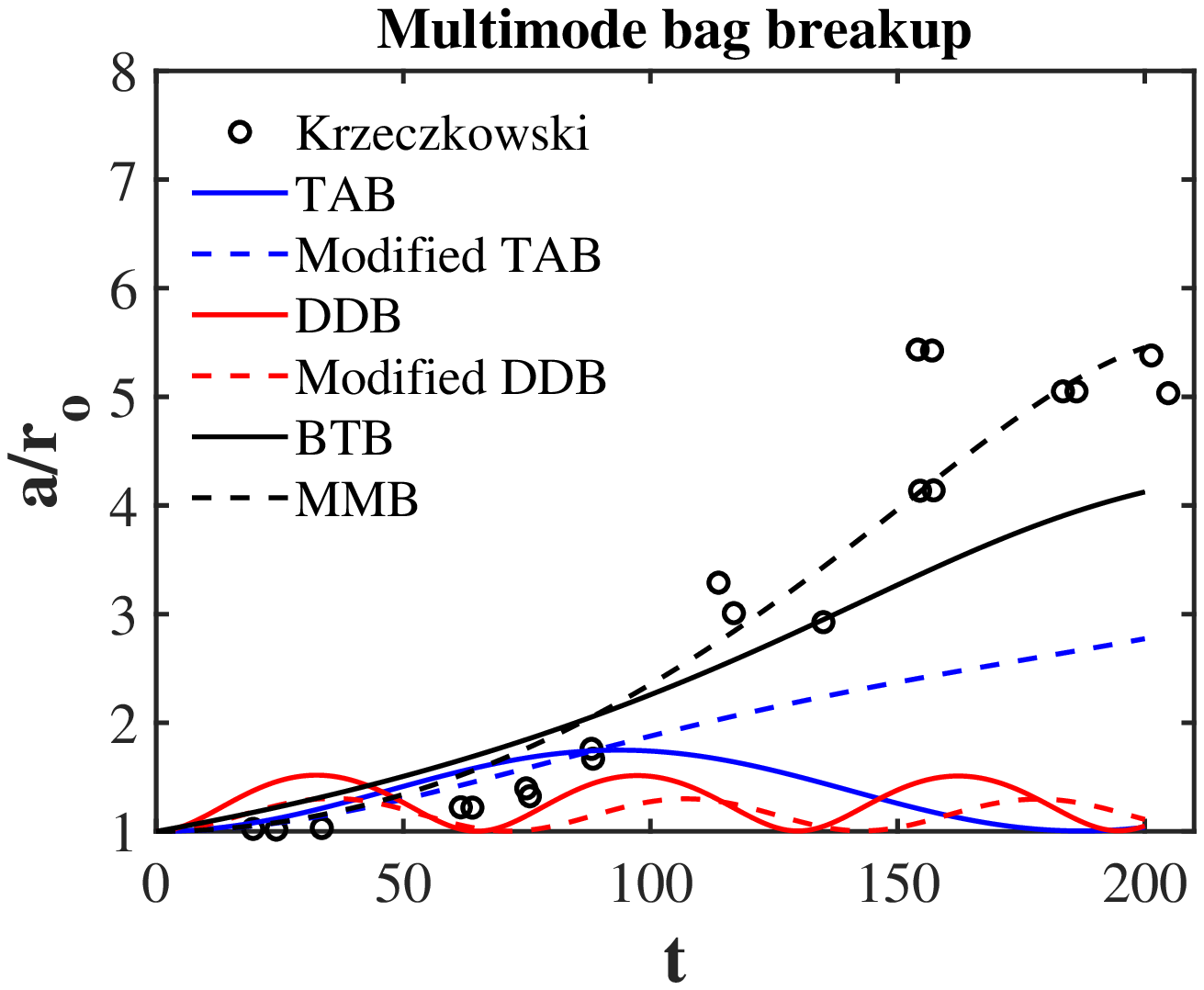}
\vspace{-0.01\textheight}
\caption{Drop deformation in multimode bag breakup regime based on different models compared to the experiment} 
\label{fig:krz_bag2}
\end{figure}

\begin{figure}[h]
\centering
\includegraphics[width=\columnwidth]{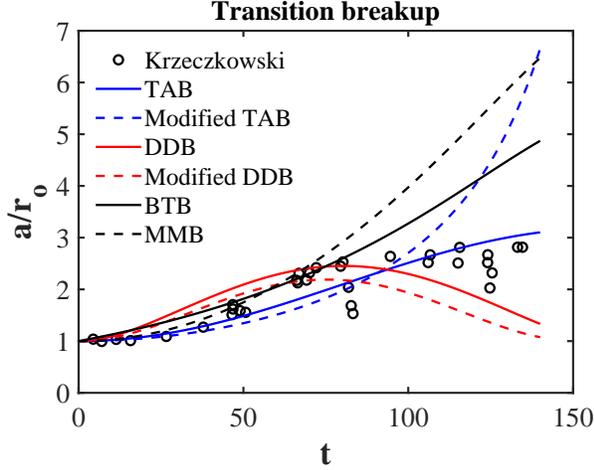}
\vspace{-0.025\textheight}
\caption{Drop deformation in transition breakup regime based on different models compared to the experiment} 
\label{fig:krz_transition}
\end{figure}

\begin{figure}[h]
\centering
\includegraphics[width=\columnwidth]{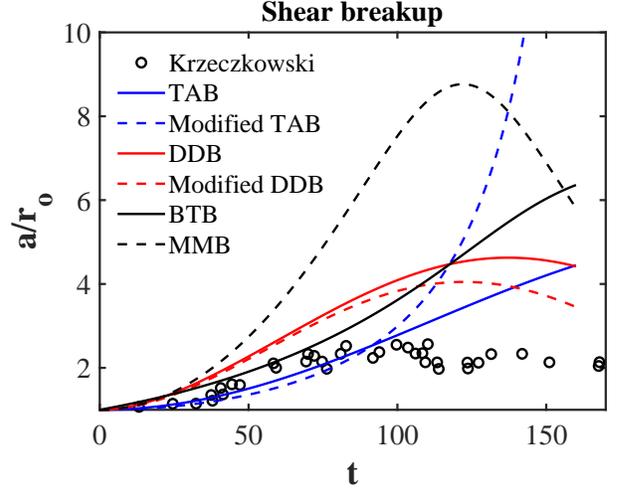}
\vspace{-0.01\textheight}
\caption{Drop deformation in shear breakup regime based on different models compared to the experiment} 
\label{fig:krz_shear}
\end{figure}

Deformation models were solved numerically using fourth order Runge-kutta method. Note that the ratio of drop diameter to its initial value, $a/r_o$, is defined differently among models. In TAB and its modified version, $a/r_o=1+0.5y$ whereas in other models $a/r_o=cy$. As shown in Fig. \ref{fig:krz_bag1} and \ref{fig:krz_bag2}, the MMB model by \cite{wang2015} predicts better deformation among others where a good agreement with experiment is achieved. TAB and DDB models and their modifications fail in predicting the large deformation involve in these cases. The modified TAB model by \cite{Park_2002} predicts the deformation better than TAB and DDB, however, it underpredicts for $t>100$. Accordingly, it can be  inferred that the MMB model developed by \cite{wang2015} would be suitable for deformation modeling of a droplet in bag and multimode bag breakup regimes. Regarding the transition regime shown in Fig. \ref{fig:krz_transition}, both TAB and DDB models show better agreement with the experiment whereas BTB, MMB and modified TAB models all together overpredict the large deformations, i.e., $t>80$. For shear-type breakup regime as plotted in Fig. \ref{fig:krz_shear}, the modified TAB model enormously over predict the experimental observation and does not follow the experimental trend. In addition, as mentioned in their work, both BTB and MMB models are suited for bag breakup regime and applying them to shear regime is naive \cite{wang2014,wang2015}. Both TAB and DDB models are within the range of experiment for shear breakup regime, however, the downward trend observed in the experiment is only captured in the DDB model and its modification by \cite{Sor_2015}. This shows that for shear breakup regime, one can employ the energy based deformation model by \cite{Ibrahim_1993}. It is worth mentioning that the robustness and predictive capability of these models would be verified better if they were compared with more experimental data in each regime. 

\begin{figure}[h]
\centering
\includegraphics[width=\columnwidth]{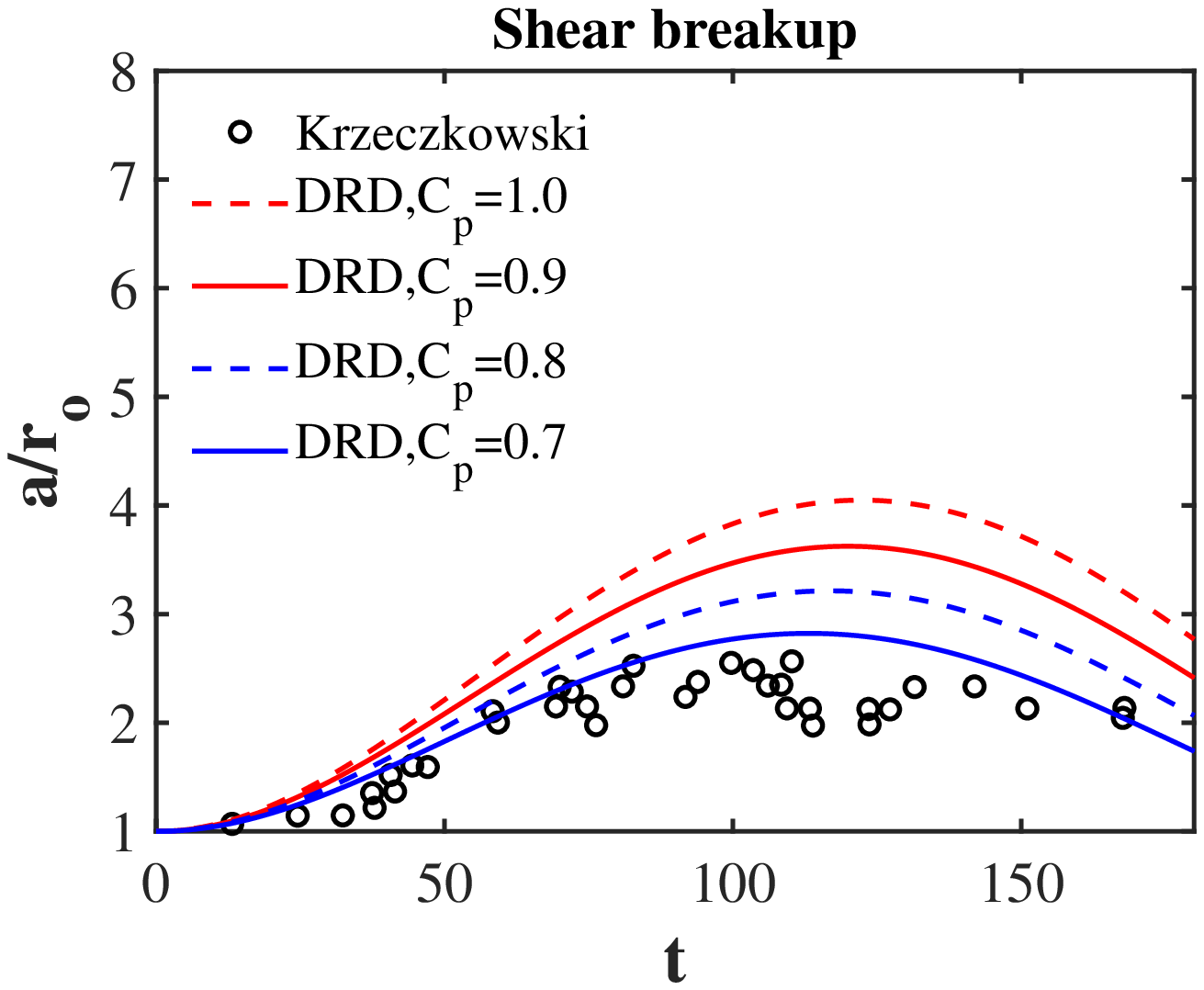}
\vspace{-0.01\textheight}
\caption{Effect of pressure coefficient on the prediction of DDB model}
\label{fig:krz_shear_cp}
\end{figure}

Moreover, as observed by \cite{Sor_2015}, the pressure term may vary during deformation and this can be accounted for by introducing a pressure coefficient, $C_p$. They found $C_p=0.93$ to better predict the corresponding experiment in the context of ice accretion, however, this may change for different flow and regimes. The effect of this parameter on the deformation of a droplet in the shear breakup regime is shown in Fig. \ref{fig:krz_shear_cp}. $c_p=0.7$ gives rise to better results for this regime revealing the fact that further modifications and tuning are required for this model and the assumption of having constant pressure on drop surface might be invalid. 

It should be mentioned that these models have to be implemented for real cases where more accurate Reynolds and Weber numbers are used rather than the constant values typically employed in the literature. However, depending on the drop relaxation and deformation time scales, one can estimate whether drop is displaced significantly during deformation. For cases where deformation occurs much faster than its displacement, then assuming a constant slip velocity during deformation would be acceptable. For a real spray case where different Weber and Reynolds numbers exist, a strategy would be required in order to switch between these models. Therefore, employing one model may result in inaccurate deformation results and its consequence effects on breakup. 

This models are intended to be tested on a case wherein series of liquid drops are injected into a cross flow and they undergo deformation before breakup occurs. As an initial test case, in order to isolate the deformation effect, a single liquid droplet is injected into a uniform flow with parameters similar to bag breakup regime. Due to the small volume loading of the drop, one-way coupling is chosen and the volumetric displacement effect for this case is conjectured to be insignificant. \cite{Hsiang_1992} observed that drag coefficient increases linearly from sphere to disk during deformation process if internal circulation is negligible. This shows that deformation has a direct influence on the dynamics of the drop through its modified drag coefficient.  \cite{Liu_etal_1993} obtained a linear relation for drag coefficient as a function of deformation parameter based on TAB model as 

\begin{equation}
    C_d = C_{d,sph}(1+2.632y)
\end{equation}

\noindent while \cite{liang2017} derived a power law relation for this coefficient as 

\begin{equation}
    C_d = 0.7y^{0.516}+0.47
\end{equation}

\begin{figure}[h]
\centering
\includegraphics[width=\columnwidth]{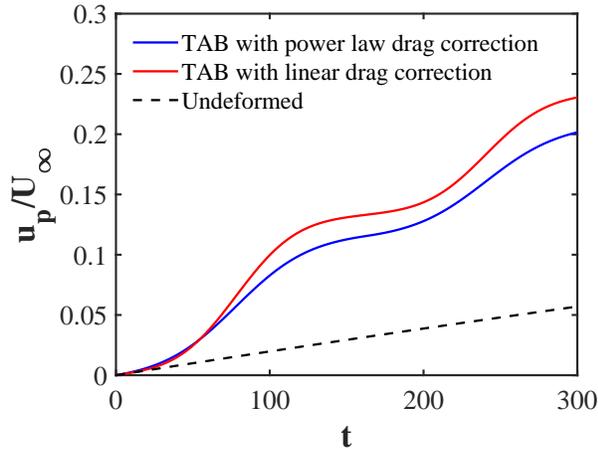}
\vspace{-0.02\textheight}
\caption{Drop deformation effect on the motion of a liquid droplet in a bag-type breakup regime}
\label{fig:u_drop_sh}
\end{figure}

In order to couple the deformation and its effect on the motion of a single droplet injected into the cross flow, cases with and without the deformation effect on the drag are compared. Bag-type breakup condition of Table\ref{tab:cases_drop} is examined before breakup occurs ($t<300$). Fig. \ref{fig:u_drop_sh} shows the results with modified drag coefficient based on the above formulations. As shown, a significant deviation is observed in the motion of droplet relative to the case where deformation is not accounted for. This could potentially alter the breakup process and affect the size and velocity of the product drops after breakup and disintegration takes place. In our future investigations, the deformation effects on the trajectory of series of liquid drops injected into the cross flow will be examined where a combination of different models will be employed to more accurately capture these effects. Then, volumetric displacement effects of deforming liquid jet into cross flow are studied in order to obtain better predictive tools for modeling dense sprays. In addition, the effect of internal circulation, which is conjectured to decrease the drag coefficient, is deemed for further investigations.

%%%%%%%%%%%%%%%%%%%%%%
\section*{Conclusions}
%%%%%%%%%%%%%%%%%%%%%%
Volumetric displacement effects of deforming liquid droplets were investigated. In order to isolate the volumtric displacement effects, deformation, coalescence, breakup and evaporation were all masked by performing a turbulent jet flow laden with a dense loading of solid particles. The standard EL two-way coupling approach was modified by accounting for the spatio-temporal variations in the local volume fraction of the carrier phase. Results of both standard and modified EL approaches were compared together to quantify the volumetric displacement effects and the regions where these effects become important. It was found that these effects increase both mean and r.m.s. velocities of the carrier phase in the region very close to the nozzle. However, they decrease further downstream due to the jet spread and dispersion of particles. Lowering the particle Stokes number increases the displacement effects on both phases. As a result, we conclude that accounting for the spatio-temporal variations in the volume fraction of the carrier phase is necessary for EL approaches of modeling dense flows. The developed model here can also be used for other configurations such as jet impingement \cite{azimi2015slot}.

In addition, in order to investigate the deformation effects, different models were tested against experimental data. It was observed that the MMB model predicts the best for bag and multimode bag breakup regimes while the original TAB agreed well with the experiment in the transition regime. The modified DDB model with a modified pressure coefficient was observed to well match the data. It was conjectured that a hybrid model based on combination of these models is required for real spray atomization flows wherein a wide range of Weber numbers and breakup regimes exists. This can also be obtained by performing fully resolved simulations \cite{azimi2018_journal,azimi2018_aps}. In order to isolate the deformation effects, as a first step, a single droplet injected into the cross flow was investigated. It was observed that accounting for deformation effect results in a significant increase in the velocity of droplet. Accordingly, we hypothesize that if volumetric coupling approach is systematically extended to the dense atomizing sprays, similar results with more pronounced effects on both displacement and deformation will be achieved.

%%%%%%%%%%%%%%%%%%%%%%%%%%
\section*{Acknowledgments}
%%%%%%%%%%%%%%%%%%%%%%%%%%
Financial support was provided under the NASA Contract Number NNX16AB07A monitored by program manager Dr. Jeff Moder, NASA Glenn Research Center. In addition, the authors acknowledge the Texas Advanced Computing Center (TACC) at The University of Texas at Austin as well as San Diego Supercomputer Center (SDSC) at University of California San Diego for providing HPC resources that have contributed to the results reported here.

\bibliographystyle{ilass}
\bibliography{ilass} 
 \end{document}